\documentclass[preprint,12pt]{elsarticle}

\usepackage{amssymb}
\usepackage{graphicx}
\usepackage{dcolumn}
\usepackage{bm}
\usepackage[utf8]{inputenc}
\usepackage{amssymb}
\usepackage{amsmath}
\usepackage{bbold}
\usepackage{pstricks}
\usepackage{color}
\usepackage{slashed}
\usepackage{braket}
\usepackage{hyperref}
\usepackage{natbib}
\usepackage{float}
\usepackage{changes}

\graphicspath{{./figures/}}

\journal{Physics Letters B}

\begin{document}

\begin{frontmatter}



\title{Schwinger pair production with ultracold atoms}


\author[HD1]{V. Kasper\corref{cor1}}
\ead{v.kasper@thphys.uni-heidelberg.de}

\author[B]{F. Hebenstreit}

\author[HD2]{M. K. Oberthaler}

\author[HD1]{J. Berges}
%\ead{j.berges@thphys.uni-heidelberg.de}

\address[HD1]{Institut f\"{u}r Theoretische Physik, Universit\"{a}t Heidelberg, Philosophenweg 16, 69120 Heidelberg, Germany}
\address[B]{Albert Einstein Center, Institut f\"{u}r Theoretische Physik, Universit\"{a}t Bern, Sidlerstrasse 5, 3012 Bern, Switzerland}
\address[HD2]{Kirchhoff Institut für Physik, Universit\"{a}t Heidelberg, Im Neuenheimer Feld 227, 69120 Heidelberg, Germany}

\cortext[cor1]{Corresponding author}

\address{}

\begin{abstract}
We consider a system of ultracold atoms in an optical lattice as a quantum simulator for electron-positron pair production in quantum electrodynamics (QED). 
For a setup in one spatial dimension, we investigate the nonequilibrium phenomenon of pair production including the backreaction leading to plasma oscillations. 
Unlike previous investigations on quantum link models, we focus on the infinite-dimensional Hilbert space of QED and show that it may be well approximated by experiments employing Bose-Einstein condensates interacting with fermionic atoms. 
Numerical calculations based on functional integral techniques give a unique access to the physical parameters required to realize  QED phenomena in a cold atom experiment. 
In particular, we use our approach to consider quantum link models in a yet unexplored parameter regime and give bounds for their ability to capture essential features of the physics. 
The results suggest a paradigmatic change towards realizations using coherent many-body states for quantum simulations of high-energy particle physics phenomena.   
\end{abstract}


\end{frontmatter}

% \linenumbers


{\it Introduction.} The creation of electron-positron pairs from the vacuum of quantum electrodynamics in an external electric field is a longstanding prediction that has not yet been directly observed~\cite{EulerHeisenberg,schwinger1951gauge}.
Upcoming experimental laser facilities, such as the Extreme Light Infrastructure (ELI) \cite{ELI}, start approaching the required critical field strength of $E_c \sim 10^{16}\,\text{V/cm}$.
Theoretically, the non-linear interplay of the produced many-body states and the applied field represents a remarkable challenge with important links to a wide range of elusive phenomena such as Unruh and Hawking radiation or string-breaking in quantum chromodynamics (QCD)~\cite{Hawking1974,Unruh1976,Bali2005}.   

Whilst the critical field strength $E_c = M^2/e$ is determined by the electron/positron mass $M$ and the absolute value of the electric charge $e$ in QED, the pair-production phenomenon essentially depends on the dimensionless ratio $E/E_c\gtrsim 1$ for an applied electric field $E$. 
In principle, physical systems with very different characteristic scales can thus be used to realize the underlying phenomenon. 
It has recently been suggested to employ a system of ultracold atoms in an optical lattice to study the physics of pair production and string breaking~\cite{Szpak2014,PhysRevLett.109.175302}.
 Even though the implementation of a gauge symmetry in an atomic setup is demanding in general~\cite{PhysRevLett.109.125302, PhysRevLett.110.125303, tagliacozzo2013optical}, it may provide a unique way of answering crucial open questions, such as regarding the nonequilibrium dynamics of the strong and electroweak sector of the standard model of particle physics probed in heavy-ion collision experiments or 
early-universe cosmology~\cite{SchmiedmayerBergesScience, CosmologicalConstant}. 

Many proposals concentrate on quantum link models~\cite{PhysRevA.88.023617} rather than QED, or even QCD. 
Since the Hilbert space of a quantum link model is finite-dimensional, the mapping to atomic systems is expected to be greatly facilitated. 
However, it is a crucial question how much of the physics of the infinite-dimensional representation corresponding to QED may be captured in practice.
Theoretical estimates based on diagonalization or matrix product states techniques are typically limited to low-dimensional representations~\cite{PhysRevLett.113.091601,kuhn2014quantum,Pichler2015}.
Recently, powerful functional integral (FI) techniques have been employed to simulate the real-time dynamics of pair production and string-breaking directly in QED on a one-dimensional lattice~\cite{PhysRevD.87.105006,2013PhRvL.111t1601H,Hebenstreit2014}, and in three dimensions~\cite{2014PhRvD..90b5016K}. 
This progress has become possible since strong bosonic fields can be efficiently sampled from coherent classical fields while keeping the full quantum nature of fermions~\cite{Aarts1998, PhysRevD.87.105006}. 

In this work, we exploit this observation and start from the infinite-dimensional representation of the QED gauge group, pointing out that it may be well approximated by experiments using Bose-Einstein condensates interacting with fermionic atoms. 
For the first time, by using FI techniques we can estimate the physical parameters required to describe the QED phenomenon of pair production in a cold atom setup and we discuss the experimental realization. For this paradigmatic example, numerical studies using the FI approach are still feasible and serve as an important benchmark for future quantum simulation experiments in numerically inaccessible regimes.
In particular, we use our approach to consider quantum link models in a yet unexplored parameter range and to give bounds for the dimensionality of the employed representation in order to capture essential features of strong-field QED in those formulations~\cite{kuhn2014quantum}.

While the FI techniques can also be applied in higher dimensions and non-Abelian gauge theories \cite{2014PhRvD..90b5016K}, we focus here on the conceptually important example of QED in one spatial dimension~\cite{Schwinger:1962tp,kogut1975hamiltonian}. 
Since there are no spatial plaquette terms in this case, angular momentum conserving atomic scattering processes can be used to directly implement the $U(1)$ gauge symmetry~\cite{PhysRevA.88.023617}. From a phenomenological point of view, this theory shares key properties with QCD as, e.g., dynamical string \mbox{breaking}, and hence provides valuable insights into nonequilibrium aspects of the theory of strong interactions.

{\it Cold atom gauge theory.} We start with the Hamiltonian formulation of lattice QED using the staggered fermion discretization~\cite{kogut1975hamiltonian}. To this end,
the spinors are decomposed such that particle and antiparticle components separately reside on two neighboring sites of the lattice. 
Introducing the operators for the fermion field $\psi_n$, the link $U_n$ being connected to the gauge potential $A_n$ and the electric field $E_n$, the Hamiltonian reads 
\begin{align}
 \label{eq:kogut_susskind}
 H_{\text{QED}}\,&=\, \sum_n \Big\{ \frac{a}{2}\, E^2_{n}+M (-1)^n \psi^{\dagger}_n \psi_n \nonumber \\
          &-\frac{i}{2a} \left[ \psi^{\dagger}_n U_n \psi_{n+1} - \psi^{\dagger}_{n+1} U_n^\dagger \psi_{n} \right]\Big\} \, ,
\end{align}
where $a$ is the lattice spacing and $g$ the gauge coupling. 
The dynamical variables of QED fulfill $[E_n,U_m] = g \delta_{nm}U_m$.
The Gauss-law operator $G_n = E_n - E_{n-1} - g \psi^{\dagger}_n \psi_n$ commutes with the Hamiltonian $[G_n,H_{\text{QED}}]=0$. 
The last relation manifests local gauge invariance. 

To make contact with quantum link models, we approximate QED by another theory, which is still gauge invariant and allows for an experimental realization with cold atoms~\cite{ANDP:ANDP201300104, PhysRevA.88.023617}.
We substitute $E_n \rightarrow g L_z$, $U_n \rightarrow [\ell(\ell+1)]^{-1/2} L_{+,n}$, where $(L_{x,n},L_{y,n},L_{z,n})$ are quantum spin operators which obey $[L_{i,n},L_{j,m}]=i\delta_{nm}\epsilon_{ijk}L_{k,n}$, and the raising operator is $ L_{+,n} = L_{x,n}+iL_{y,n}$. 
The performed substitution renders the dimension of the local Hilbert space finite-dimensional.
Consequently, the QED relation $[U_n,U_m^{\dagger}]= 0$ is no longer valid but is replaced by $[L_{+,n}, L_{-,m}] = 2 \delta_{nm} L_{z,m}$. Remarkably, local gauge invariance is not affected.

The Schwinger representation of the angular momentum operators allows us to express the quantum spins by bosonic degrees of freedom $b_n$ and $d_n$~\cite{assa1994interacting}:
$L_{+,n} = b_n^{\dagger}d_n$, $L_{-,n} = d_n^{\dagger}b_n$ and $L_{z,n}=(b^{\dagger}_n b_n - d^{\dagger}_n d_n)/2$, with the constraint $b^{\dagger}_n b_n + d^{\dagger}_{n} d_n = 2 \ell$.
Here, $\ell$ denotes the spin magnitude fixing the number of bosonic atoms. 
In this representation, the  Hamiltonian describing the cold 
atom (CA) system  becomes
\begin{align}
\!\!\! H_{\text{CA}} &= \sum_n \!\Big\{ \frac{ g^2a}{4} [  b^{\dagger}_n b^{\dagger}_n b_n b_n + d^{\dagger}_n d^{\dagger}_n d_n d_n ] + M (-1)^n \psi^{\dagger}_n \psi_n \notag \\
           &-\frac{i}{ 2 a \sqrt{\ell(\ell + 1)}} \! \left[\psi^{\dagger}_n b_n^{\dagger} d_n \psi_{n+1} - \psi^{\dagger}_{n+1} d_n^\dagger b_n\psi_{n}\right]\!\Big\}\!\!
\label{eq:HColdAtomQED}
\end{align}
and 
depends on two species of bosonic operators $b_n, d_n$ living on links and fermionic operators $\psi_n$ located on lattice sites.
The parameters are determined by the basic physical quantities of the cold atom system, such as the gauge coupling $g$ given by the on-site scattering processes of the bosons and the spin-changing collisions between the fermionic and bosonic atoms. 
 
Employing the density-phase representation with $b_n = \sqrt{\ell + \delta \rho_{b,n}} e^{i\theta_{b,n}}$ and $d_n = \sqrt{\ell + \delta \rho_{d,n}} e^{i\theta_{d,n}}$, one finds $H_{\text{CA}} = H_{\text{QED}} + \mathcal{O}(\delta \rho / \ell)$.
Therefore, the number of bosonic atoms per site, $2\ell$, controls the approximation 
and allows to tune from the quantum link formulation to lattice QED by increasing $\ell$. 

In previous work, the Hamiltonian \eqref{eq:HColdAtomQED} was studied for $\ell \sim \mathcal{O}(1)$ via diagonalization or matrix product states methods~\cite{PhysRevLett.109.175302,PhysRevLett.113.091601,Banuls:2013jaa,kuhn2014quantum,Pichler2015}. 
Here we consider for the first time the dynamics in the regime $\ell \gg 1$ to approach lattice QED. From an experimental point of view, this regime corresponds to putting Bose-Einstein condensates on the links rather than single bosonic atoms.

{\it Functional integral approach.} 
To study the strong-field regime of QED, the field strength needs to be of the order of the critical field  $E_c=M^2/g$.
The corresponding cold atom setup is characterized by $E_c=g|N_b - N_d|/2 \sim M^2/g$, where $N_b,N_d$ denote the number of atoms in the Bose-Einstein condensates. 
For $N_b,N_d \sim \mathcal{O}(\ell)\gg 1$, the FI approach of Refs.~\cite{Aarts1998,PruschkeGelfand,PhysRevD.87.105006,2014PhRvD..90b5016K,Hebenstreit2014} allows us to study the dynamics in this regime. 

To this end, we denote the bosonic fields collectively by $\phi_n = \begin{pmatrix} b^{\dagger}_{n}, b_{n}, d^{\dagger}_{n}, d_n \end{pmatrix}$ and define the generating functional for correlation functions in the presence of sources $J_n = \begin{pmatrix} J_{b,n},  J^{\ast}_{b,n},  J_{d,n},  J^{\ast}_{d,n} \end{pmatrix}$ by $Z[J] = \operatorname{Tr} \{ \rho_0 T_{\mathcal{C}} e^{i J \cdot \phi}\}$. 
Here, $\rho_0$ is the initial density matrix, $J \cdot \phi = \sum_n \int_t J_n(t) \cdot \phi_n(t)$ with $t$ the time coordinate along the closed time path~$\mathcal{C}$, and $T_{\mathcal{C}}$ denotes time-ordering along~$\mathcal{C}$.
Employing the coherent state basis, the matrix element of the density operator at initial time is $\bra{+} \rho_0 \ket{-}$, where $\ket{+}$ and $\ket{-}$ are the first coherent states on the forward and backward contour, respectively.
The FI representation of the generating functional is
\begin{align}
 Z[J] = \int [d\phi] [d \psi^{\dagger} d \psi] \bra{+} \rho_0 \ket{-} e^{iS + iJ\cdot \phi} \, 
\end{align}
with the action
\begin{align}
 S = \int_t \sum_n ( \psi_n^{\dagger} i \partial_t \psi_n +b_n^{\dagger} i \partial_t b_n  + d_n^{\dagger} i \partial_t d_n ) - H_{\text{CA}} \, .
\end{align}
We analytically perform the Gaussian integral for the fermions and then expand to first order in the bosonic response field $\tilde{\phi}_n$, which arises from the Keldysh rotation $\phi_n = \bar{\phi}_n + \operatorname{sgn}_{\mathcal{C}}\tilde{\phi}_n$.
Disregarding higher-order terms, i.e.\ neglecting subleading corrections in bosonic occupancies~\cite{2014PhRvD..90b5016K} that are suppressed by $N_{b}^{-1}, N_{d}^{-1} \ll 1$, leads to the self-consistent set of equations 
\begin{align}
i \partial_t b_n &= \frac{g^2 a }{2} b^{\dagger}_n b_n b_n +i\frac{ d_n F_{n+1\,n}}{4a\sqrt{\ell(\ell+1)}} \ , \notag \\
i \partial_t d_n &= \frac{ g^2 a}{2} d^{\dagger}_n d_n d_n -i\frac{ b_n F_{n\,n+1} }{4a \sqrt{\ell(\ell+1)}} \ , \notag \\ 
i \partial_t F_{nm} &= \sum_{n'}{ \left[h^{\text{CA}}_{nn'} F_{n'm} - F_{nn'} h^{\text{CA}}_{n'm}\right]} \label{eq:CAEom}\ .
\end{align}
Here, $F_{nm} = \braket{[\psi_n, \psi^{\dagger}_m]}$ is the fermion equal-time correlation function, whose evolution is governed by 
\begin{align}
 h^{\text{CA}}_{nm} &= \frac{i [d^{\dagger}_{n-1} b_{n-1}  \delta_{n-1\, m} - b^{\dagger}_n d_n \delta_{n+1\, m} ]}{ 2 a \sqrt{\ell(\ell+1)} } + M (-1)^n \delta_{n\,m} . \notag
 %\label{eq:HCAFermions}
\end{align}
The equations \eqref{eq:CAEom} preserve the Gauss law, $\partial_t G_n=0$, if initialized accordingly. 
A similar derivation for the QED Hamiltonian \eqref{eq:kogut_susskind} gives
\begin{align}
\partial_t E_n &= \frac{g}{2a} \operatorname{Re} [F_{n+1\,n}U_n]  , \notag \\
\partial_t U_n &= i g a E_n U_n \ , \notag \\
i \partial_t F_{nm} &= \sum_{n'} [h^{\text{QED}}_{nn'} F_{n'm} - F_{nn'} h^{\text{QED}}_{n'm}]  \ , \label{eq:KSEom}
\end{align} 
with
\begin{align}
 h^{\text{QED}}_{nm} = \frac{i}{2a}[U^{\ast}_{n-1}\delta_{n-1\,m}-U_n \delta_{n+1\,m}] + M (-1)^n \delta_{n\,m} \, . \notag
 %\label{eq:HKSFermions}
\end{align}
In fact, by taking the time derivative of $E_n\to g (b^{\dagger}_n b_n - d^{\dagger}_n d_n)/2$ and $U_n\to b^{\dagger}_n d_n$ and inserting the density-phase representation, one finds that \eqref{eq:CAEom} approximates \eqref{eq:KSEom} with a truncation error of $\mathcal{O}(\delta \rho/\ell)$.
In the following, we keep $\ell$ as a parameter to answer the question whether essential properties of QED can be captured for finite $\ell$. 

\begin{figure}[t]
 \center{
 \includegraphics[width=0.7\columnwidth]{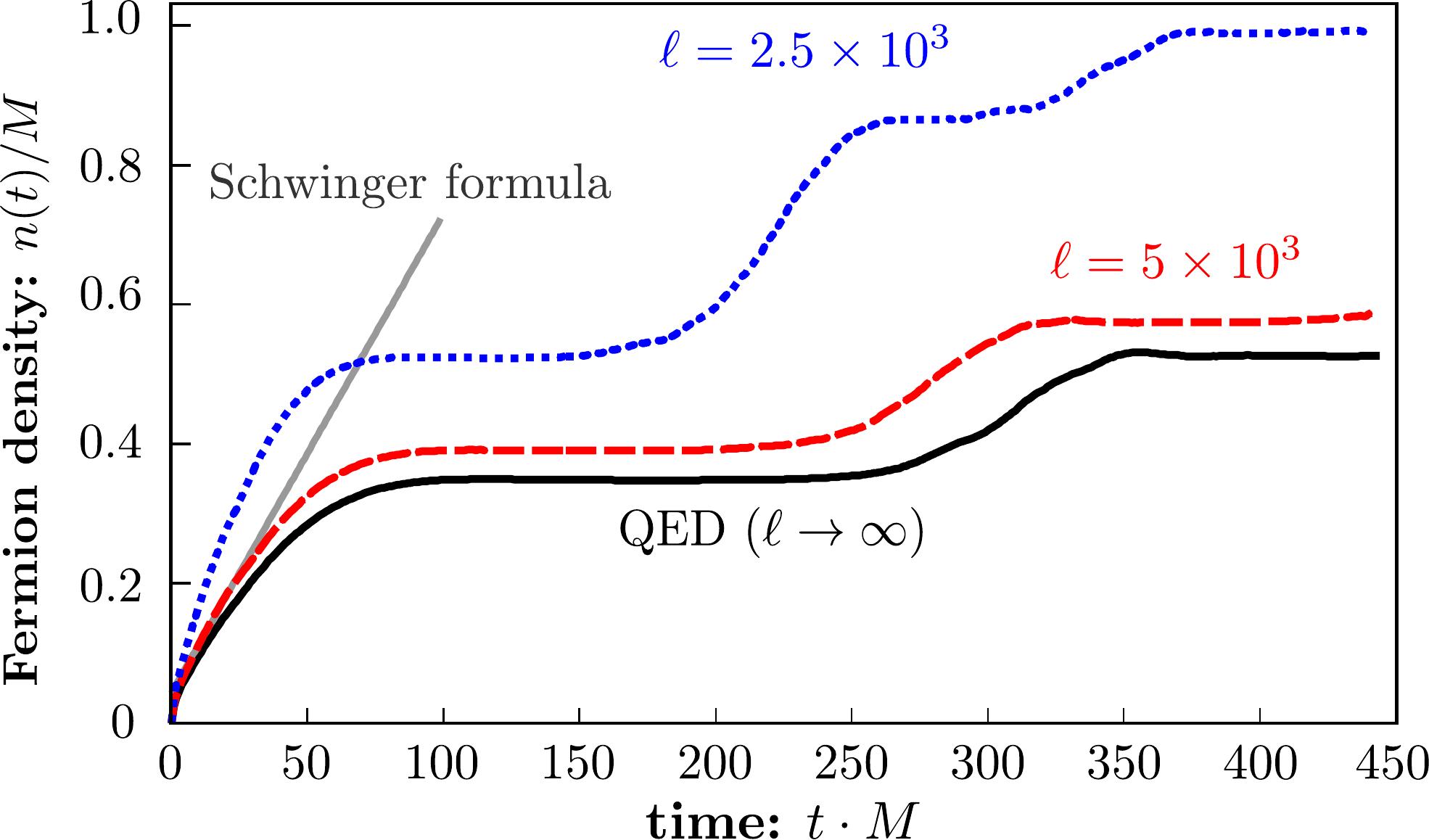}}
 \caption{Time evolution of the fermion number density in the cold atom system for different numbers of atoms $\sim \ell$ as compared to the QED result ($\ell \rightarrow \infty$). The straight line corresponds to the Schwinger formula, which neglects the backreaction of the produced fermions on the applied field.}
\label{fig:1}
\end{figure}

{\it Pair production.} The creation of electron-positron pairs in a uniform electric field $E$ may be viewed as a quantum process in which virtual electron-positron dipoles can be separated to become real pairs once they gain the binding energy of twice the rest mass energy.
This QED process has been estimated~\cite{EulerHeisenberg,schwinger1951gauge} neglecting the backreaction of the produced pairs on the applied field, and the analytic result for the rate $\dot{n}= M^2 E/(2\pi E_c) \exp (-\pi E_c/ E)$ is depicted in Fig.~\ref{fig:1}.

This estimate should be valid at sufficiently early times and provides an important benchmark for any simulation method. 
Therefore, we consider a spatial lattice of length $N a$ with periodic boundary conditions and leave more refined estimates taking into account specific trap geometries of cold atom systems for further studies. 
We first compute the real-time evolution according to \eqref{eq:CAEom} for $g/M=0.1$ and $aM=0.005$ in the limit $\ell \rightarrow \infty$, where it agrees to QED described by \eqref{eq:KSEom}. 
Here $N$ determines the number of fermionic atoms, and we checked that for the largest employed lattices with $N=512$ no significant volume dependence can be observed and our results are insensitive to changes in the lattice spacing. 
Accordingly, employing a standard definition of the particle number density~\cite{PhysRevD.87.105006,2014PhRvD..90b5016K} the simulation result for QED ($\ell = \infty$) as shown in Fig.~\ref{fig:1} agrees well with the Schwinger formula at early times.
At later times the backreaction of the produced pairs on the applied field is seen to give the expected sizable corrections~\cite{PhysRevD.87.105006,2013PhRvL.111t1601H}. 
  
\begin{figure}[t]
 \center{
 \includegraphics[width=0.7\columnwidth]{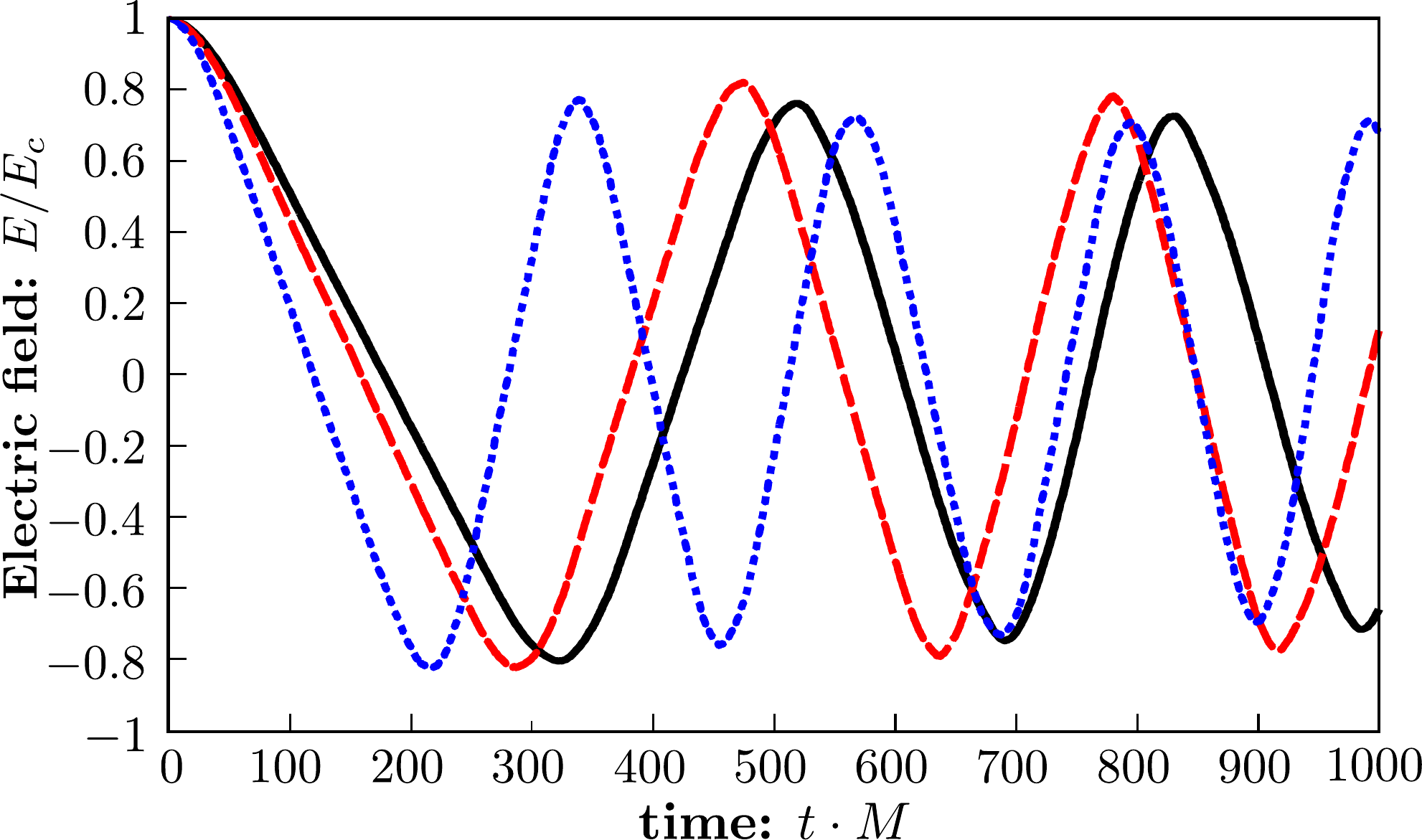}}
 \caption{Dynamics of the homogeneous  electric field as represented by the bosonic species population imbalance of the atomic system for different values of $\ell$ as in Fig.~\ref{fig:1}. 
 The plasma oscillations arise from the backreaction of the produced pairs on the applied field.}
 \label{fig:2}
\end{figure}
  
Of course, in the corresponding cold atom system no particles are produced as the number of atoms is fixed. 
However, since two neighboring fermions are considered as particles and antiparticles in the staggered formulation, pair production is encoded in the hopping of atoms between odd and even sites of the optical lattice.
This generates correlations, whose time evolution describe the corresponding phenomenon of pair production as shown in Fig.~\ref{fig:1}.
The results demonstrate the convergence of the atomic system's dynamics to the QED behavior as the number of bosonic atoms is increased. 
For $\ell=2500$ we still observe considerable deviations from the QED result, whereas the difference becomes small for $\ell=5000$.

In Fig.~\ref{fig:2} the time evolution of the cold atom analogue of the electric field, $E= g (N_b - N_d)/2$, is given for different values of $\ell$. 
As for Fig.~\ref{fig:1}, we start with a bosonic species imbalance $N_b-N_d=2M^2/g^2>0$ corresponding to the critical electric field strength in QED, and the analogue of the Dirac vacuum or `Fermi sea' with the lowest $N/2$ energy eigenstates occupied. 
By comparison to Fig.~\ref{fig:1}, we observe a decrease of the electric field as the fermion number increases due to pair production. 
The correlated hopping of fermions reduces the bosonic species imbalance until it becomes zero and even changes its sign with $N_b-N_d<0$ giving rise to plasma oscillations~\cite{PhysRevD.45.4659,2014PhRvD..90b5016K,PhysRevD.87.105006}. 
At times when the corresponding electric field drops below a critical level, particle creation effectively terminates as reflected in the characteristic plateaus in the particle number density. 

\begin{figure}[t]
  \center{
 \includegraphics[width=0.7\columnwidth]{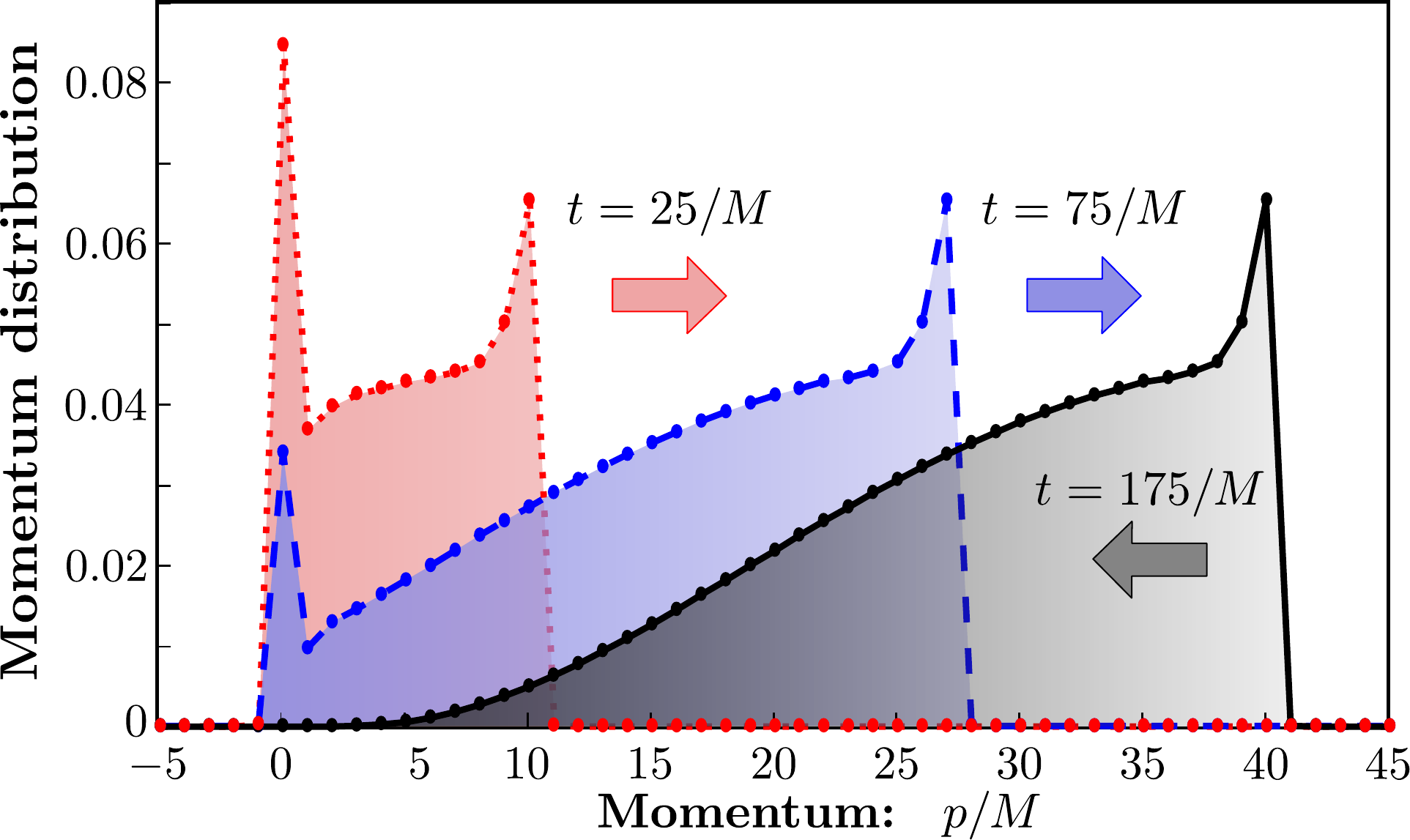}}
 \caption{Momentum distribution of the produced fermions at $tM= 25$ ({\it dotted}), $tM=75$ ({\it dashed}) and $tM=175$ ({\it solid}) for $\ell = 10^4$ such that the QED result is well described.}
\label{fig:3}
\end{figure}

The plasma oscillations are caused by the production and subsequent acceleration of particle-antiparticle pairs, which can be observed from their momentum distribution~\cite{PhysRevD.87.105006,2014PhRvD..90b5016K} as shown in Fig.~\ref{fig:3} for $\ell = 10^4$ such that the QED result is well reproduced.
The homogeneous electric field dominantly produces fermions around zero momentum, and accelerates the particles to higher momenta during the subsequent evolution.
As the electric field decreases due to energy conservation, the production amplitude around zero momentum drops as well.
At a time $tM\sim175$, the fermions reach their maximum momentum along with a vanishing net electric field. 
Subsequently, the fermionic current results in a further decrease of the electric field to negative values along with a deceleration of the produced particles leading to the observed plasma oscillations.

\begin{figure}[t!]
\center{
\includegraphics[width=0.7\columnwidth]{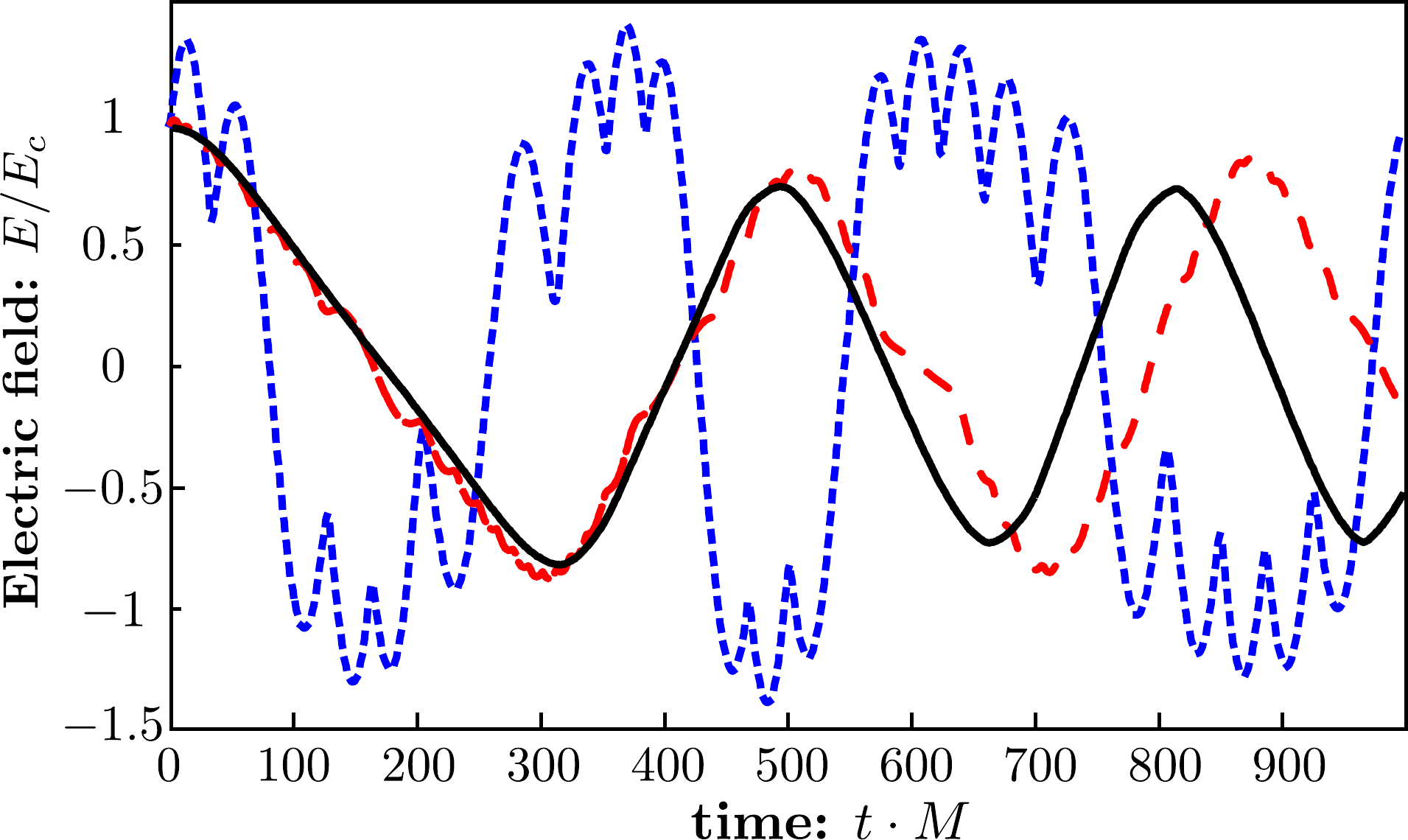}
}
\caption{Time evolution of the electric field for different total numbers of fermionic atoms, $N=32$ ({\it dotted}), $N=128$ ({\it dashed}) and $N=512$ ({\it solid}) with $\ell=10^4$. }

\label{fig:4}

\end{figure}

The limiting experimental resources enforce a study of the dependence on the total number $N$ of fermionic atoms. Fig.~\ref{fig:4} shows the electric field for different $N$ with fixed $\ell=10^4$.  We observe a reasonable description of the QED results for a full oscillation period employing a total number of fermionic atoms down to about $N = 128$, with sizable deviations occurring at later times. For $N = 512$ accurate descriptions are achieved for the entire range of times we considered.   

{\it Experimental realization.} The physics of QED pair production in one spatial dimension may already be realized with available experimental techniques~\cite{PhysRevLett.109.175302, PhysRevA.88.023617,2011PhRvL.107A5301Z, MorschOberthaler}. Here we point out how the
relevant regime of large $\ell$ may be efficiently implemented and manipulated experimentally with the help of  coherent many-body states. 

Regarding the bosonic degrees of freedom, we confine two substates of one hyperfine manifold of bosonic atoms in a one-dimensional geometry via an external potential.
Further, we employ a red detuned laser to generate an optical lattice such that the atoms are localized and the nearest neighbor hopping is suppressed. 
Already this construction allows us to realize mesoscopic bosonic gases with two components ($b_n$ and $d_n$) per site which can be described by the one axis twisting Hamiltonian ~\cite{Muessel} corresponding to the first two terms of \eqref{eq:HColdAtomQED}.
The preparation of the bosonic atoms mimicking the electric field can be achieved by a magnetic field or homogeneous two-photon microwave coupling, which produces a coherent spin state between the two components with a non-vanishing  population difference.

Concerning the fermionic degrees of freedom, we again trap two substates of one hyperfine state manifold of fermionic atoms in a one-dimensional geometry. 
The aforementioned optical lattice is blue detuned for the fermions.
Consequently, the fermions are located between the bosonic links, which is an essential ingredient of \eqref{eq:HColdAtomQED}.
In addition, we superimpose a second optical lattice with double lattice spacing in order to generate the staggered structure of the fermions.
The frequency of this laser is tuned  closer to resonance with respect to the fermions than to the bosons, such that the second lattice does not affect the latter ones.
The staggered structure leads to a mini-gap in the dispersion relation of the fermions such that the initial state of the fermions corresponds to a fully filled lowest mini-band~\cite{Koehl}.
The detection of the fermions can be achieved by exploiting the band mapping technique~\cite{Scelle2013}.

The overlap of neighboring bosonic and fermionic atoms makes the hopping of one fermion from one site to the next possible via spin changing collisions. This modifies the internal state of the boson, i.e., species $b$ become $d$ or vice versa.
This interaction induced hopping process  implemented by  boson-fermion spin-exchange collisions locally preserves the total spin~\cite{Li}. 
In particular, the detuning of this process that is experimentally controlled by the external field corresponds to the mass term in \eqref{eq:HColdAtomQED}.
The dynamics of the system is initiated with a quench of the mass term from being far off-resonant.

{\it Conclusion.} Simulating high-energy physics by experiments with ultracold atoms may be achieved with coherent many-body states rather than single atoms.
Our findings of the required resources in terms of atom numbers and protocols may already be realized with available experimental techniques. 
This opens new possibilities to resolve questions in the strong-coupling regimes of gauge theories, where no alternative real-time simulation techniques are known so far.
Together with recent experimental proofs of concept in strongly interacting systems~\cite{SchmiedmayerBerges2015}, one may hope to realize the old dream of solving complex problems in quantum field theory by experiment.\\

We thank M.~Dalmonte, E.~Demler, A.~Frishman, T.~Gasenzer, J.~G\"oltz,
F. Jendrzejewski, M.~Karl, N.~M\"uller, J.~Pawlowski, A.~Polkovnikov, I.~Rocca and U.-J.~Wiese for helpful discussions and collaborations on related work. 
V.~Kasper is supported by the Max Planck Society.
F.~Hebenstreit acknowledges support from the Alexander von Humboldt Foundation in the early stages of this work as well as from the European Research Council under the European Union's Seventh Framework Programme (FP7/2007-2013)/ ERC grant agreement 339220. This work is part of and supported by the
DFG Collaborative Research Centre "SFB 1215 (ISOQUANT)".



%% References


\bibliographystyle{model1-num-names}
\biboptions{sort&compress}
\bibliography{Papers}

\begin{thebibliography}{36}
\expandafter\ifx\csname natexlab\endcsname\relax\def\natexlab#1{#1}\fi
\providecommand{\bibinfo}[2]{#2}
\ifx\xfnm\relax \def\xfnm[#1]{\unskip,\space#1}\fi
%Type = Article
\bibitem[{{Heisenberg} and {Euler}(1936)}]{EulerHeisenberg}
\bibinfo{author}{W.~{Heisenberg}}, \bibinfo{author}{H.~{Euler}},
\newblock \bibinfo{title}{{Folgerungen aus der Diracschen Theorie des
  Positrons}},
\newblock \bibinfo{journal}{Z. Phys.} \bibinfo{volume}{98}
  (\bibinfo{year}{1936}) \bibinfo{pages}{714--732}.
%Type = Article
\bibitem[{Schwinger(1951)}]{schwinger1951gauge}
\bibinfo{author}{J.~Schwinger},
\newblock \bibinfo{title}{On gauge invariance and vacuum polarization},
\newblock \bibinfo{journal}{Phys. Rev.} \bibinfo{volume}{82}
  (\bibinfo{year}{1951}) \bibinfo{pages}{664}.
%Type = Article
\bibitem[{{ELI delivery consortium}(2016)}]{ELI}
\bibinfo{author}{{ELI delivery consortium}},
\newblock \bibinfo{journal}{http://www.eli-laser.eu/}  (\bibinfo{year}{2016}).
%Type = Article
\bibitem[{{Hawking}(1974)}]{Hawking1974}
\bibinfo{author}{S.~W. {Hawking}},
\newblock \bibinfo{title}{{Black hole explosions?}},
\newblock \bibinfo{journal}{Nature} \bibinfo{volume}{248}
  (\bibinfo{year}{1974}) \bibinfo{pages}{30--31}.
%Type = Article
\bibitem[{{Unruh}(1976)}]{Unruh1976}
\bibinfo{author}{W.~G. {Unruh}},
\newblock \bibinfo{title}{{Notes on black-hole evaporation}},
\newblock \bibinfo{journal}{Phys. Rev. D} \bibinfo{volume}{14}
  (\bibinfo{year}{1976}) \bibinfo{pages}{870--892}.
%Type = Article
\bibitem[{{Bali} et~al.(2005){Bali}, {Neff}, {D{\"u}ssel}, {Lippert}, and
  {Schilling}}]{Bali2005}
\bibinfo{author}{G.~S. {Bali}}, \bibinfo{author}{H.~{Neff}},
  \bibinfo{author}{T.~{D{\"u}ssel}}, \bibinfo{author}{T.~{Lippert}},
  \bibinfo{author}{K.~{Schilling}},
\newblock \bibinfo{title}{{Observation of string breaking in QCD}},
\newblock \bibinfo{journal}{Phys. Rev. D} \bibinfo{volume}{71}
  (\bibinfo{year}{2005}) \bibinfo{pages}{114513}.
%Type = Article
\bibitem[{{Szpak} and {Sch{\"u}tzhold}(2012)}]{Szpak2014}
\bibinfo{author}{N.~{Szpak}}, \bibinfo{author}{R.~{Sch{\"u}tzhold}},
\newblock \bibinfo{title}{{Optical lattice quantum simulator for quantum
  electrodynamics in strong external fields: spontaneous pair creation and the
  Sauter-Schwinger effect}},
\newblock \bibinfo{journal}{New J. Phys} \bibinfo{volume}{14}
  (\bibinfo{year}{2012}) \bibinfo{pages}{035001}.
%Type = Article
\bibitem[{Banerjee et~al.(2012)Banerjee, Dalmonte, M\"uller, Rico, Stebler,
  Wiese, and Zoller}]{PhysRevLett.109.175302}
\bibinfo{author}{D.~Banerjee}, \bibinfo{author}{M.~Dalmonte},
  \bibinfo{author}{M.~M\"uller}, \bibinfo{author}{E.~Rico},
  \bibinfo{author}{P.~Stebler}, \bibinfo{author}{U.-J. Wiese},
  \bibinfo{author}{P.~Zoller},
\newblock \bibinfo{title}{Atomic quantum simulation of dynamical gauge fields
  coupled to fermionic matter: From string breaking to evolution after a
  quench},
\newblock \bibinfo{journal}{Phys. Rev. Lett.} \bibinfo{volume}{109}
  (\bibinfo{year}{2012}) \bibinfo{pages}{175302}.
%Type = Article
\bibitem[{Zohar et~al.(2012)Zohar, Cirac, and Reznik}]{PhysRevLett.109.125302}
\bibinfo{author}{E.~Zohar}, \bibinfo{author}{J.~I. Cirac},
  \bibinfo{author}{B.~Reznik},
\newblock \bibinfo{title}{Simulating compact quantum electrodynamics with
  ultracold atoms: Probing confinement and nonperturbative effects},
\newblock \bibinfo{journal}{Phys. Rev. Lett.} \bibinfo{volume}{109}
  (\bibinfo{year}{2012}) \bibinfo{pages}{125302}.
%Type = Article
\bibitem[{Banerjee et~al.(2013)Banerjee, B\"ogli, Dalmonte, Rico, Stebler,
  Wiese, and Zoller}]{PhysRevLett.110.125303}
\bibinfo{author}{D.~Banerjee}, \bibinfo{author}{M.~B\"ogli},
  \bibinfo{author}{M.~Dalmonte}, \bibinfo{author}{E.~Rico},
  \bibinfo{author}{P.~Stebler}, \bibinfo{author}{U.-J. Wiese},
  \bibinfo{author}{P.~Zoller},
\newblock \bibinfo{title}{Atomic quantum simulation of $\mathbf{U}(n)$ and
  $\mathrm{SU}(n)$ non-abelian lattice gauge theories},
\newblock \bibinfo{journal}{Phys. Rev. Lett.} \bibinfo{volume}{110}
  (\bibinfo{year}{2013}) \bibinfo{pages}{125303}.
%Type = Article
\bibitem[{Tagliacozzo et~al.(2013)Tagliacozzo, Celi, Zamora, and
  Lewenstein}]{tagliacozzo2013optical}
\bibinfo{author}{L.~Tagliacozzo}, \bibinfo{author}{A.~Celi},
  \bibinfo{author}{A.~Zamora}, \bibinfo{author}{M.~Lewenstein},
\newblock \bibinfo{title}{Optical abelian lattice gauge theories},
\newblock \bibinfo{journal}{Ann. Phys.} \bibinfo{volume}{330}
  (\bibinfo{year}{2013}) \bibinfo{pages}{160--191}.
%Type = Article
\bibitem[{{Schmiedmayer} and {Berges}(2013)}]{SchmiedmayerBergesScience}
\bibinfo{author}{J.~{Schmiedmayer}}, \bibinfo{author}{J.~{Berges}},
\newblock \bibinfo{title}{{Cold Atom Cosmology}},
\newblock \bibinfo{journal}{Science} \bibinfo{volume}{341}
  (\bibinfo{year}{2013}) \bibinfo{pages}{1188--1189}.
%Type = Article
\bibitem[{{Finazzi} et~al.(2012){Finazzi}, {Liberati}, and
  {Sindoni}}]{CosmologicalConstant}
\bibinfo{author}{S.~{Finazzi}}, \bibinfo{author}{S.~{Liberati}},
  \bibinfo{author}{L.~{Sindoni}},
\newblock \bibinfo{title}{{Cosmological Constant: A Lesson from Bose-Einstein
  Condensates}},
\newblock \bibinfo{journal}{Phys. Rev. Lett.} \bibinfo{volume}{108}
  (\bibinfo{year}{2012}) \bibinfo{pages}{071101}.
%Type = Article
\bibitem[{Zohar et~al.(2013)Zohar, Cirac, and Reznik}]{PhysRevA.88.023617}
\bibinfo{author}{E.~Zohar}, \bibinfo{author}{J.~I. Cirac},
  \bibinfo{author}{B.~Reznik},
\newblock \bibinfo{title}{Quantum simulations of gauge theories with ultracold
  atoms: Local gauge invariance from angular-momentum conservation},
\newblock \bibinfo{journal}{Phys. Rev. A} \bibinfo{volume}{88}
  (\bibinfo{year}{2013}) \bibinfo{pages}{023617}.
%Type = Article
\bibitem[{Buyens et~al.(2014)Buyens, Haegeman, Van~Acoleyen, Verschelde, and
  Verstraete}]{PhysRevLett.113.091601}
\bibinfo{author}{B.~Buyens}, \bibinfo{author}{J.~Haegeman},
  \bibinfo{author}{K.~Van~Acoleyen}, \bibinfo{author}{H.~Verschelde},
  \bibinfo{author}{F.~Verstraete},
\newblock \bibinfo{title}{Matrix product states for gauge field theories},
\newblock \bibinfo{journal}{Phys. Rev. Lett.} \bibinfo{volume}{113}
  (\bibinfo{year}{2014}) \bibinfo{pages}{091601}.
%Type = Article
\bibitem[{K{\"u}hn et~al.(2014)K{\"u}hn, Cirac, and
  Ba{\~n}uls}]{kuhn2014quantum}
\bibinfo{author}{S.~K{\"u}hn}, \bibinfo{author}{J.~I. Cirac},
  \bibinfo{author}{M.~Ba{\~n}uls},
\newblock \bibinfo{title}{Quantum simulation of the schwinger model: A study of
  feasibility},
\newblock \bibinfo{journal}{Phys. Rev. A} \bibinfo{volume}{90}
  (\bibinfo{year}{2014}) \bibinfo{pages}{042305}.
%Type = Article
\bibitem[{{Pichler} et~al.(2015){Pichler}, {Dalmonte}, {Rico}, {Zoller}, and
  {Montangero}}]{Pichler2015}
\bibinfo{author}{T.~{Pichler}}, \bibinfo{author}{M.~{Dalmonte}},
  \bibinfo{author}{E.~{Rico}}, \bibinfo{author}{P.~{Zoller}},
  \bibinfo{author}{S.~{Montangero}},
\newblock \bibinfo{title}{{Real-time Dynamics in U(1) Lattice Gauge Theories
  with Tensor Networks}}  (\bibinfo{year}{2015}).
%Type = Article
\bibitem[{Hebenstreit et~al.(2013)Hebenstreit, Berges, and
  Gelfand}]{PhysRevD.87.105006}
\bibinfo{author}{F.~Hebenstreit}, \bibinfo{author}{J.~Berges},
  \bibinfo{author}{D.~Gelfand},
\newblock \bibinfo{title}{Simulating fermion production in $1\mathbf{+}1$
  dimensional qed},
\newblock \bibinfo{journal}{Phys. Rev. D} \bibinfo{volume}{87}
  (\bibinfo{year}{2013}) \bibinfo{pages}{105006}.
%Type = Article
\bibitem[{{Hebenstreit} et~al.(2013){Hebenstreit}, {Berges}, and
  {Gelfand}}]{2013PhRvL.111t1601H}
\bibinfo{author}{F.~{Hebenstreit}}, \bibinfo{author}{J.~{Berges}},
  \bibinfo{author}{D.~{Gelfand}},
\newblock \bibinfo{title}{{Real-Time Dynamics of String Breaking}},
\newblock \bibinfo{journal}{Phys. Rev. Lett.} \bibinfo{volume}{111}
  (\bibinfo{year}{2013}) \bibinfo{pages}{201601}.
%Type = Article
\bibitem[{{Hebenstreit} and {Berges}(2014)}]{Hebenstreit2014}
\bibinfo{author}{F.~{Hebenstreit}}, \bibinfo{author}{J.~{Berges}},
\newblock \bibinfo{title}{{Connecting real-time properties of the massless
  Schwinger model to the massive case}},
\newblock \bibinfo{journal}{Phys. Rev. D} \bibinfo{volume}{90}
  (\bibinfo{year}{2014}) \bibinfo{pages}{045034}.
%Type = Article
\bibitem[{{Kasper} et~al.(2014){Kasper}, {Hebenstreit}, and
  {Berges}}]{2014PhRvD..90b5016K}
\bibinfo{author}{V.~{Kasper}}, \bibinfo{author}{F.~{Hebenstreit}},
  \bibinfo{author}{J.~{Berges}},
\newblock \bibinfo{title}{{Fermion production from real-time lattice gauge
  theory in the classical-statistical regime}},
\newblock \bibinfo{journal}{Phys. Rev. D} \bibinfo{volume}{90}
  (\bibinfo{year}{2014}) \bibinfo{pages}{025016}.
%Type = Article
\bibitem[{{Aarts} and {Smit}(1999)}]{Aarts1998}
\bibinfo{author}{G.~{Aarts}}, \bibinfo{author}{J.~{Smit}},
\newblock \bibinfo{title}{{Real-time dynamics with fermions on a lattice}},
\newblock \bibinfo{journal}{Nucl. Phys. B} \bibinfo{volume}{555}
  (\bibinfo{year}{1999}) \bibinfo{pages}{355--394}.
%Type = Article
\bibitem[{Schwinger(1962)}]{Schwinger:1962tp}
\bibinfo{author}{J.~Schwinger},
\newblock \bibinfo{title}{{Gauge Invariance and Mass. 2.}},
\newblock \bibinfo{journal}{Phys. Rev.} \bibinfo{volume}{128}
  (\bibinfo{year}{1962}) \bibinfo{pages}{2425--2429}.
%Type = Article
\bibitem[{Kogut and Susskind(1975)}]{kogut1975hamiltonian}
\bibinfo{author}{J.~Kogut}, \bibinfo{author}{L.~Susskind},
\newblock \bibinfo{title}{Hamiltonian formulation of wilson's lattice gauge
  theories},
\newblock \bibinfo{journal}{Phys. Rev. D} \bibinfo{volume}{11}
  (\bibinfo{year}{1975}) \bibinfo{pages}{395}.
%Type = Article
\bibitem[{Wiese(2013)}]{ANDP:ANDP201300104}
\bibinfo{author}{U.-J. Wiese},
\newblock \bibinfo{title}{Ultracold quantum gases and lattice systems: quantum
  simulation of lattice gauge theories},
\newblock \bibinfo{journal}{Ann. Phys.} \bibinfo{volume}{525}
  (\bibinfo{year}{2013}) \bibinfo{pages}{777--796}.
%Type = Book
\bibitem[{Auerbach(1994)}]{assa1994interacting}
\bibinfo{author}{A.~Auerbach}, \bibinfo{title}{Interacting electrons and
  quantum magnetism}, \bibinfo{publisher}{Springer Science \& Business Media,
  New York}, \bibinfo{year}{1994}.
%Type = Article
\bibitem[{Bañuls et~al.(2013)Bañuls, Cichy, Cirac, and
  Jansen}]{Banuls:2013jaa}
\bibinfo{author}{M.~Bañuls}, \bibinfo{author}{K.~Cichy},
  \bibinfo{author}{J.~I. Cirac}, \bibinfo{author}{K.~Jansen},
\newblock \bibinfo{title}{The mass spectrum of the schwinger model with matrix
  product states},
\newblock \bibinfo{journal}{JHEP} \bibinfo{volume}{2013}
  (\bibinfo{year}{2013}).
%Type = Article
\bibitem[{{Berges} et~al.(2011){Berges}, {Gelfand}, and
  {Pruschke}}]{PruschkeGelfand}
\bibinfo{author}{J.~{Berges}}, \bibinfo{author}{D.~{Gelfand}},
  \bibinfo{author}{J.~{Pruschke}},
\newblock \bibinfo{title}{{Quantum Theory of Fermion Production after
  Inflation}},
\newblock \bibinfo{journal}{Phys. Rev. Lett.} \bibinfo{volume}{107}
  (\bibinfo{year}{2011}) \bibinfo{pages}{061301}.
%Type = Article
\bibitem[{Kluger et~al.(1992)Kluger, Eisenberg, Svetitsky, Cooper, and
  Mottola}]{PhysRevD.45.4659}
\bibinfo{author}{Y.~Kluger}, \bibinfo{author}{J.~M. Eisenberg},
  \bibinfo{author}{B.~Svetitsky}, \bibinfo{author}{F.~Cooper},
  \bibinfo{author}{E.~Mottola},
\newblock \bibinfo{title}{Fermion pair production in a strong electric field},
\newblock \bibinfo{journal}{Phys. Rev. D} \bibinfo{volume}{45}
  (\bibinfo{year}{1992}) \bibinfo{pages}{4659--4671}.
%Type = Article
\bibitem[{{Zohar} and {Reznik}(2011)}]{2011PhRvL.107A5301Z}
\bibinfo{author}{E.~{Zohar}}, \bibinfo{author}{B.~{Reznik}},
\newblock \bibinfo{title}{{Confinement and Lattice Quantum-Electrodynamic
  Electric Flux Tubes Simulated with Ultracold Atoms}},
\newblock \bibinfo{journal}{Phys. Rev. Lett.} \bibinfo{volume}{107}
  (\bibinfo{year}{2011}) \bibinfo{pages}{275301}.
%Type = Article
\bibitem[{{Morsch} and {Oberthaler}(2006)}]{MorschOberthaler}
\bibinfo{author}{O.~{Morsch}}, \bibinfo{author}{M.~{Oberthaler}},
\newblock \bibinfo{title}{{Dynamics of Bose-Einstein condensates in optical
  lattices}},
\newblock \bibinfo{journal}{Rev. Mod. Phys.} \bibinfo{volume}{78}
  (\bibinfo{year}{2006}) \bibinfo{pages}{179--215}.
%Type = Article
\bibitem[{{Muessel} et~al.(2014){Muessel}, {Strobel}, {Linnemann}, {Hume}, and
  {Oberthaler}}]{Muessel}
\bibinfo{author}{W.~{Muessel}}, \bibinfo{author}{H.~{Strobel}},
  \bibinfo{author}{D.~{Linnemann}}, \bibinfo{author}{D.~B. {Hume}},
  \bibinfo{author}{M.~K. {Oberthaler}},
\newblock \bibinfo{title}{{Scalable Spin Squeezing for Quantum-Enhanced
  Magnetometry with Bose-Einstein Condensates}},
\newblock \bibinfo{journal}{Phys. Rev. Lett.} \bibinfo{volume}{113}
  (\bibinfo{year}{2014}) \bibinfo{pages}{103004}.
%Type = Article
\bibitem[{K\"ohl et~al.(2005)K\"ohl, Moritz, St\"oferle, G\"unter, and
  Esslinger}]{Koehl}
\bibinfo{author}{M.~K\"ohl}, \bibinfo{author}{H.~Moritz},
  \bibinfo{author}{T.~St\"oferle}, \bibinfo{author}{K.~G\"unter},
  \bibinfo{author}{T.~Esslinger},
\newblock \bibinfo{title}{{Fermionic Atoms in a Three Dimensional Optical
  Lattice: Observing Fermi Surfaces, Dynamics, and Interactions}},
\newblock \bibinfo{journal}{Phys. Rev. Lett.} \bibinfo{volume}{94}
  (\bibinfo{year}{2005}) \bibinfo{pages}{080403}.
%Type = Article
\bibitem[{{Scelle} et~al.(2013){Scelle}, {Rentrop}, {Trautmann}, {Schuster},
  and {Oberthaler}}]{Scelle2013}
\bibinfo{author}{R.~{Scelle}}, \bibinfo{author}{T.~{Rentrop}},
  \bibinfo{author}{A.~{Trautmann}}, \bibinfo{author}{T.~{Schuster}},
  \bibinfo{author}{M.~K. {Oberthaler}},
\newblock \bibinfo{title}{{Motional Coherence of Fermions Immersed in a Bose
  Gas}},
\newblock \bibinfo{journal}{Phys. Rev. Lett.} \bibinfo{volume}{111}
  (\bibinfo{year}{2013}) \bibinfo{pages}{070401}.
%Type = Article
\bibitem[{{Li} et~al.(2015){Li}, {Zhu}, {He}, {Wang}, {Guo}, {Xu}, {Zhang}, and
  {Wang}}]{Li}
\bibinfo{author}{X.~{Li}}, \bibinfo{author}{B.~{Zhu}},
  \bibinfo{author}{X.~{He}}, \bibinfo{author}{F.~{Wang}},
  \bibinfo{author}{M.~{Guo}}, \bibinfo{author}{Z.-F. {Xu}},
  \bibinfo{author}{S.~{Zhang}}, \bibinfo{author}{D.~{Wang}},
\newblock \bibinfo{title}{{Coherent Heteronuclear Spin Dynamics in an Ultracold
  Spinor Mixture}},
\newblock \bibinfo{journal}{Phys. Rev. Lett.} \bibinfo{volume}{114}
  (\bibinfo{year}{2015}) \bibinfo{pages}{255301}.
%Type = Article
\bibitem[{{Schweigler} et~al.(2015){Schweigler}, {Kasper}, {Erne}, {Rauer},
  {Langen}, {Gasenzer}, {Berges}, and {Schmiedmayer}}]{SchmiedmayerBerges2015}
\bibinfo{author}{T.~{Schweigler}}, \bibinfo{author}{V.~{Kasper}},
  \bibinfo{author}{S.~{Erne}}, \bibinfo{author}{B.~{Rauer}},
  \bibinfo{author}{T.~{Langen}}, \bibinfo{author}{T.~{Gasenzer}},
  \bibinfo{author}{J.~{Berges}}, \bibinfo{author}{J.~{Schmiedmayer}},
\newblock \bibinfo{title}{{On solving the quantum many-body problem}}
  \bibinfo{volume}{arXiv:1505.03126} (\bibinfo{year}{2015}).

\end{thebibliography}






% \end{thebibliography}

\end{document}